\documentstyle[12pt]{article}

\newcommand{\be}{\begin{eqnarray}}
\newcommand{\en}{\end{eqnarray}}
\newcommand{\ph}{\phantom{00}}
\newcommand{\no}{\nonumber}
\newcommand{\dsl}{\mbox{$\partial$}\hspace*{-2,0mm}\mbox{/}\hspace*{0,5mm}}

\def\biggg#1{{\hbox{$\left#1\vbox to 20.5pt{}\right.\n@space$}}}

\def\Biggg#1{{\hbox{$\left#1\vbox to 23.5pt{}\right.\n@space$}}}

\title{                                                                         
{\vspace{-3cm} \normalsize                                                      
\hfill \parbox{30mm}{DESY 96-083}}\\[25mm]                                  
Simulating the massive Schwinger model \\
with chiral defect fermions   \\[8mm]} 
\author{                                                                        
A. Jaster \\[4mm] 
Deutsches Elektronen-Synchrotron DESY, \\                          
Notkestr.\,85, D-22603 Hamburg, Germany}                                        
                                                                                
\date{April, 1996}

\begin{document}

\maketitle
\begin{abstract}

Some time ago  Kaplan proposed a new model 
for the description of chiral fermions on the lattice
by adding an extra dimension for the fermions.
A variant of this proposal was introduced by Shamir and can be used to
describe vector-like 
theories in even dimensions. We used this model 
for the simulation of the massive
Schwinger model at different gauge couplings. The prediction that the 
fermion mass
gets only multiplicative renormalization was tested and verified.
 
\end{abstract}


\section{Introduction}

Some years ago
Kaplan suggested a new method simulating chiral fermions on the
lattice \cite{Kaplan}. He started with a vector-like odd-dimensional 
theory and
a mass term that depends on one space
coordinate and has the form of a domain wall. He found a zero-mode bound to 
this
domain wall. From the point of view of the even-dimensional domain wall, this
zero-mode is a chiral fermion. The zero-mode exists up to a critical 
momentum, so that the chirality is a low energy effect. Anomalies in 
gauge currents arise from the flow of charge into the extra 
dimension. In the case of a finite extra dimension which has the topology
of a circle and periodic boundary conditions
an additional anti-domain wall with a chiral fermion of
opposite handness is generated.

A similar but technicaly simpler approach was made by Shamir \cite{Shamir}.
Instead of an infinite extra dimension and a mass that is a function of this
dimension,
he used a semi-infinite coordinate with free boundary conditions
and a constant odd-dimensional mass.
In the case of a finite lattice this corresponds to cutting the links at both
domains in the original model and dropping the side with a negative mass. 
Therefore, in numerical simulations one has to take only
half as many points to obtain similar results. The zero-states with different
chirality are located at the boundaries.  A connection of these two boundaries
by a link of strength $m$
will form a Dirac fermion from  the two Weyl fermions  whose mass is
proportional to $m$, if the exponentially small overlap between the 
surface states can be neglected. This model can be used to describe 
vector-like theories in even dimensions.
Unlike the case of chiral gauge theories 
\cite{Golterman} in the vector-like model 
there is no problem to couple 
($2n+1$)-dimensional  fermions to a gauge
field in $2n$ dimensions.  
It was shown that perturbative corrections to the quark mass   
are proportional to $m$, so that it undergoes 
only multiplicative
renormalization to all orders \cite{Shamir}. 
Therefore, there is no fine tuning problem that appears for Wilson fermions.
This model has in the massless case a very mild breaking of the axial
symmetries.
Moreover, the correct chiral limit can be proven \cite{FURSHA}. 

We examined the case of $(2+1)$-dimensional fermions coupled to a
two-dimensional $U(1)$ gauge field, i.e.\ the massive Schwinger model
\cite{SCHWINGER,COJASU}.
To test the prediction that the fermion mass gets only 
multiplicative
renormalization  we studied the behaviour of the pseudoscalar
isotriplet ('pion') mass if $m$ tends to
zero.


\section{Free boundary fermions}

We start our discussion with the free fermion case in $2+1$ dimensions and a
lattice of size $\Omega=L^2 \cdot L_{\rm s}$.
The action is given by
\be
S_{\rm F}&=&\sum_x \Biggl \{
\sum_{s=1}^{L_{\rm s}} 
\overline{\Psi}_{x,s} \left [ \dsl + M + \frac{r}{2} \Delta 
\right ] \Psi _{x,s}  
\label{FreeAction}
+   m   \left [ \overline{\Psi}_{x,1} P_{\rm L}
\Psi_{x,L_{\rm s}} + \overline{\Psi}_{x,L_{\rm s}}
P_{\rm R} \Psi_{x,1}  \right ] \Biggr \} \ ,
\en
where $\dsl=\sum_{\mu=1}^{3} \sigma_{\mu} \partial_{\mu}$ and 
$\Delta=\sum_{\mu=1}^{3} \Delta_{\mu} $.
$\sigma_{\mu}$ are the Pauli matrices, $r$ denotes the Wilson
parameter and $P_{\rm R,L}= (1 \pm \sigma_3)/2$  are the projection
operators.
A lattice point is given by $(x,s)$, where $s$ labels the
extra coordinate. In the following pages you will note that
the lattice spacing is taken to be $a=1$.
The lattice derivation $\partial_{\mu}$
and Laplacian $\Delta_{\mu}$   for $\mu=1,2$ are defined as usual and
\be
\partial_3 \Psi _{x,s} \equiv \frac{1}{2} \left \{ \begin{array}{ll}
\Psi_{x,2} & s=1 \\
\Psi_{x,s+1}-\Psi_{x,s-1} & 2\leq s \leq L_{\rm s}-1 \\
-\Psi_{x,L_{\rm s}-1} & s=L_{\rm s} \\  
\end{array} 
\right. \ ,
\en
\be
\Delta_3 \Psi _{x,s} \equiv  \left \{ \begin{array}{ll}
\Psi_{x,2}-2\Psi_{x,1} & s=1 \\
\Psi_{x,s+1}-2\Psi_{x,s}+\Psi_{x,s-1} & 2\leq s \leq L_{\rm s}-1 \\
-2\Psi_{x,L_{\rm s}}+\Psi_{x,L_{\rm s}-1} & s=L_{\rm s} \\  
\end{array} 
\right. \ ,
\en
i.e.\ the case $m=0$ corresponds to open boundaries in the third direction.
For $r=1$ and $m=1$ the model corresponds to the topology of a circle and  
periodic boundary condition in $s$, for $r=1$ and $m=-1$ it 
describes  the antiperiodic case. Thus, in these
cases the (2+1)-dimensional boundary fermion model is equivalent to a
three-dimensional model with Wilson fermions and the corresponding boundary
conditions.

\begin{figure}                        
\vspace{8.0cm}                                                                  
\includegraphics{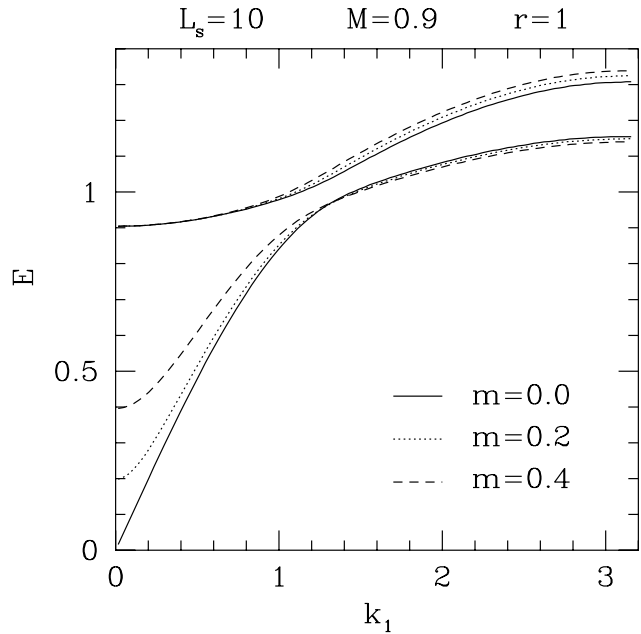}                                              
\begin{center}                                                                  
\parbox{15cm}{\caption{\label{fig_ImpulsAbh}                        
Part of the energy spectrum for a lattice extension of $L_{\rm s}=10$
and a $(2+1)$-dimensional mass $M=0.9$. The three different types of lines
characterize the different values of the mass parameter $m$.
}}                                                                              
\end{center}                                                                    
\end{figure}

The action (\ref{FreeAction}) 
leeds to the Hamiltonian
\be
\label{ShamirM}
H=\sigma_2 \left [ {\rm i}\, \sigma_1 \overline{k}_1
+ \sigma_3 \partial_3 + M + \frac{r}{2} \left (
- {{\hat{k}_1} }^2 + \Delta_3 \right )
+ m \left ( \delta_{s,1} \delta_{s',L_{\rm s}} P_{\rm L}
          + \delta_{s,L_{\rm s}} \delta_{s',1} P_{\rm R}  \right )
\right ] \ ,
\en
\be
\overline{k}=\sin(k) \ , \ph\ph \hat{k}=2 \sin(k/2) \ . \no
\en 
Its eigenvalues occur in pairs corresponding to particle and anti-particle
and were already examined for $m=0$ \cite{Jansen,Horvath}. In this case the
chiral zero-modes are confined to the two boundaries. The wave functions have
the form of plane waves in the two-dimensional space and decay 
exponentially with $(1-M)$ in
$s$. The chiral zero-modes exist for momenta $k_1$ below some critical 
momentum
$k_{\rm c}$. For $0<M/r<2$ there is one zero-mode at each boundary. 
For $m\neq 0$, an
infinite lattice in $x_1$ 
direction and a finite third direction one can calculate the lowest 
eigenvalues in the limit where $m^2$ and ${k_1}^2$ are small, i.e.\
$m^2$, ${k_1}^2 \ll M^2$. If one choose $L_{\rm s}$ large enough, so that the
exponentially small mixing of the two chiral modes can be neglected
($(1-M)^{L_{\rm s}}  \ll m$), the result is 
\be
{E_0}^2 = {k_1}^2+m^2 M^2 (2-M)^2 \ .
\en
Therefore, the mass of the light Dirac fermion is 
\be
\overline{m}=mM(2-M) \ .
\en
In Fig.\ \ref{fig_ImpulsAbh} we sketch the dependence of the energy spectrum 
for the two lightest states on the mass. One can observe
that the lowest eigenvalue has indeed the predicted form.
 
\section{Interacting theory}

The interacting theory is defined by coupling the fermions to a
two-dimensional gauge field, i.e.\ $U_{(x,s)\mu} = U_{x\mu}$ for 
$\mu=1, \, 2$ and $U_{(x,s)3}= 1$. This leads for $r=1$ to the following
gauged fermionic part of the action 
\be
S_{\rm F} 
&\equiv& \left ( \overline{\Psi}, Q[U] \Psi  \right ) \no \\
\no \\
&=& \sum_x \sum_{s=1}^{L_{\rm s}} \Biggl \{ \frac{1}{2}
\sum_{\mu=1}^{2} \left [ \overline{\Psi}_{x,s} \left( 1 +
\sigma_\mu \right) U_{x \mu} \Psi_{x+\hat{\mu},s} +
\overline{\Psi}_{x+\hat{\mu},s}  \left ( 1 - \sigma_\mu \right )
U_{x \mu}^* \Psi_{x,s} \right ] 
+ \left( M - 3 \right )
\overline{\Psi}_{x,s} \Psi_{x,s}  \Biggr \} \no \\
&& + \sum_x \Biggl \{ \sum_{s=1}^{L_{\rm s}-1} \left [ \overline{\Psi}_{x,s}
P_{\rm R} \Psi_{x,s+1} + \overline{\Psi}_{x,s+1} P_{\rm L}
\Psi_{x,s}  \right ] 
+ m  \left [ \overline{\Psi}_{x,L_{\rm s}} P_{\rm R} \Psi_{x,1}
+ \overline{\Psi}_{x,1} P_{\rm L} \Psi_{x,L_{\rm s}}   \right ] \Biggr \} \ ,
\en
where $U_{x\mu}$ denotes the U(1) gauge variable. Apart from the
unconventional sign of the mass term, the first term is the usual
two-dimensional action. The preferred boundary conditions are periodic in 
space direction and antiperiodic in time direction.  The gauge part is 
given by \be
S_{\rm G}= \beta \sum_P \left ( 1 - {\rm Re \,} U_P \right ) \ ,
\en
where $U_P$ is the product of four $U$'s around a plaquette and $\beta = 
1/g^2$. The whole action is given by the sum of the two parts:
\be
S=S_{\rm F}+S_{\rm G} \ .
\en
The case $m=0$ should yield to the massless Schwinger model by taking 
first the limit $L_{\rm s} \rightarrow \infty$ and then $\beta 
\rightarrow \infty$. 

The definitions of the vector and axial currents as 
well as their divergence equations can be found in \cite{SHAMIR2,FURSHA}.
Here we just give a definition of the axial transformation (for even 
$L_{\rm s}$):
\be
\delta_A^i \, \psi_{x,s} &=& {\rm i} \, q(s) \,  \sigma_i  \, \psi_{x,s}, \\
\delta_A^i \, \overline{\psi}_{x,s} &=& -{\rm i}  \, q(s) \,  
\overline{\psi}_{x,s}  \, \sigma_i,
\en
where
\be
q(s) \equiv  \left \{ \begin{array}{ll}
1 & 1 \le s \le L_{\rm s}/2 \\
-1 & L_{\rm s}/2 < s \le L_{\rm s} \\
\end{array} 
\right. \ .
\en
The non-invariance of the action under this transformation results from 
the coupling between the layers 
$s=L_{\rm s}/2$ and $s=L_{\rm s}/2+1$, respectively, 
$s=1$ and $s=L_{\rm s}$ for $m \ne 0$.

For the numerical simulations we used the Hybrid Monte Carlo algorithm
with pseudo\-fermions
\cite{DKPR}. This requires a positive definite determinant and therefore the
flavour duplication of the fermion spectrum. The fermion matrix for the first
species ist $Q[U]$, for the second $Q[U]^\dagger$. 
If we define the reflection ${\cal R}$ by
\be
{\cal R} \Psi_{x,s} \equiv \Psi_{x,L_{\rm s}+1-s} \ ,
\en
the fermion matrix satisfies
\be
\sigma_3 {\cal R} \, Q[U] \, \sigma_3 {\cal R} = Q[U]^\dagger \ .
\en
This yields to 
\be
\det Q[U] = \det Q[U]^\dagger \ .
\en
Thus, the simulated model
describes the two flavour massive Schwinger model.

The topological charge is given by 
\be
Q_{\rm top}= \frac{1}{2\pi} \sum_x F_x \ , \ph\ph Q_{\rm top} \in
Z\!\!\!Z \ ,
\en
\be
\exp \left ( {\rm i}\, F_x  \right ) = U_{x,1} \, 
U_{x+\hat{1},2} \,  U_{x+\hat{2},1}^* \,  U_{x,2}^* \ , \ph\ph 
F_x \in [ -\pi , \pi ) \ .
\en
The different topological sectors are separated by a potential barrier 
of height $2 \beta$, so that the tunneling is suppressed. We restricted the 
simulations to the topological trivial case 
($Q_{\rm top}=0$)\footnote{Contributions of the non-trivial sectors 
were examined in \cite{DILGER}},
i.e.\ we start with $U_{x\mu}=1$ and controlled that the topological 
charge doesn't change. The 
lightest particles in this model are a pseudoscalar isotriplet with mass
$M_{1^-}$ satisfying
\be
\frac{M_{1^-}}{g} 
\mathrel{\mathop {- \!\!\! -  \!\!\! 
\rightarrow}_{\frac{\overline{m}}{g} \rightarrow 0}} 
c \left ( \frac{\overline{m}}{g} \right ) 
^\frac{2}{3} \ , 
\en
\be
c=6 \, \sqrt{\frac{2}{\pi}} \left ( 
\frac{e^\gamma}{2 \pi} \right )^\frac{2}{3} \approx 2.07 \ .
\en
This is the analogue to the pion in QCD and 
becomes massless as the fermion mass $\overline{m}$ goes to zero.
The matrix inversions were performed by 
the conjugate gradient algorithm \cite{JASTER}.

It can be shown that in the limit $L_{\rm s} \rightarrow \infty$ 
corrections to the fermion mass $\overline{m}$ are proportional $m$, so that 
the fermion mass  
in the boundary model gets only multiplicative renormalization.
The reason is that in weak coupling perturbation theory, corrections to the
inverse fermion propagator are exponentially suppressed with $|s-s'|$.
Therefore, the zero-modes can't mix, except for an exponentially small factor,
and a mass term can't develop.
Thus, the chiral limit is attained if $m$ goes to zero.
We used the massive Schwinger model with two degenerate fermion species to
test non-pertubatively
the prediction that the fermion mass tends to zero for $m \rightarrow 0$.

\begin{figure}                        
\vspace{8.5cm}                                                                  
\includegraphics{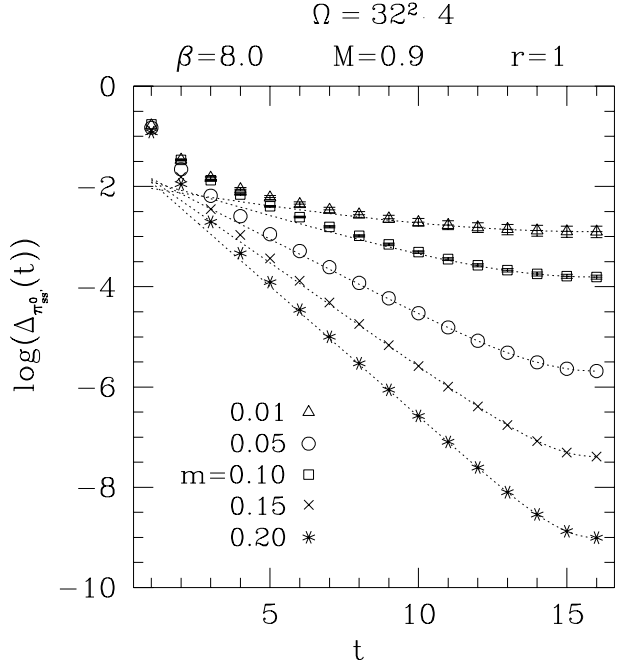}                                              
\begin{center}                                                                  
\parbox{15cm}{\caption{\label{KSF4.2}                                       
Correlation function for the  pion for $(s,s')=(1,L_{\rm s})$. The error 
bars for $m\ge 0.1$ were omitted
because, they are too small for a visualization (smaller than 3\% of the 
measured values). The curves shown are the best fit in the interval 
$13 \le t \le 16$. 
}}                                                                              
\end{center}                                                                    
\end{figure}                                                                    
The simulations were performed at $M=0.9$, so that at tree level one has 
$\overline{m}=0.99 \, m$. For the gauge coupling 
the two values $\beta=0.25$
and $\beta=8.0$ were examined. In the last case the renormalization of
$M$ is small. It follows that 
the exponentially decay in the additional direction is
fast and the exponentially contribution to the mass are even for small
values of $L_{\rm s}$ strongly suppressed. Therefore, only small extensions of 
 $L_{\rm s}$ are needed. As a check we used lattices of
size $\Omega=L_{\rm x}\cdot L_{\rm t}\cdot L_{\rm s}=12\cdot 24\cdot 10$
and $\Omega=32^2\cdot 4$. For the mass parameter $m$ we have chosen the
values $m=0.0, \, 0.1, \dots , \, 0.5$
in the first case and $m=0.01, \, 0.05, \, 0.10, \, 0.15, \, 0.20$
in the second case. At $\beta=0.25$ we examined $m=0.0, \, 0.1, \, 0.2, 
\, 0.3$ for a $12\cdot 24\cdot 10$ lattice.

\begin{figure}[t]                        
\vspace{7.5cm}                                                                  
\includegraphics{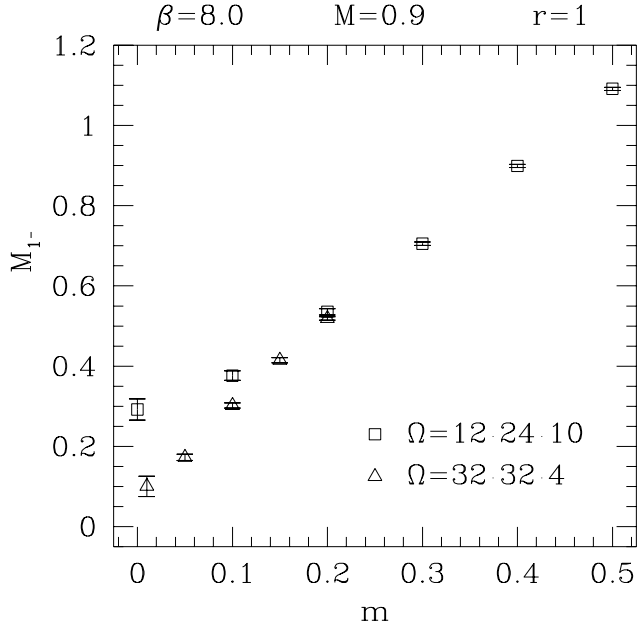}                                              
\begin{center}                                                                  
\parbox{15cm}{\caption{\label{KSF5.2}   
Pion mass as a function of   $m$ at $\beta=8.0$. 
}}                                                                
\end{center}                                                                    
\end{figure}                                                                    
\begin{figure}[t]                        
\vspace{8.0cm}                                                                  
\includegraphics{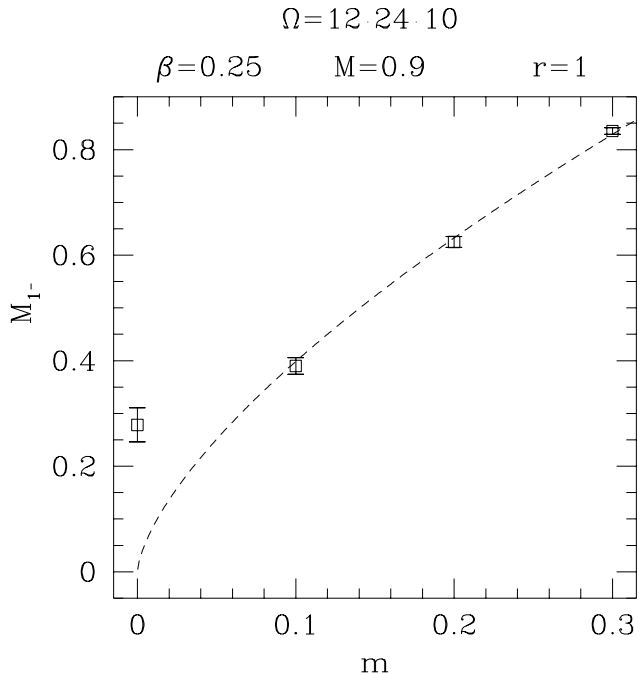}                                              
\begin{center}                                                                  
\parbox{15cm}{\caption{\label{KSF6.2}   
Pion mass as a function of  $m$ at $\beta=0.25$. 
The dashed line indicates an ansatz proportional $m^{2/3}$.
}}                                                                
\end{center}                                                                    
\end{figure}                                                                    
The pion mass $M_{1^-}$ is calculated from the correlation function
\be
\Delta_{\pi^0_{ss'}}(t) = \left \langle  \pi^0_{ss'}(0)
\pi^0_{ss'}(t)  \right \rangle \ ,
\en
\be
\pi^0_{ss'}(t) \equiv  \sum_{\rm x} 
\overline{\psi}_{({\rm x},{\rm t}),s} \sigma_3 \tau_3 \psi_{({\rm 
x},{\rm t}),s'} \ , 
\en
where $(I,I_3)^P=(1,0)^-$ and $x=({\rm x},{\rm t})$. 
For $(s,s')=(1,L_{\rm s})$ and 
$(s,s')=(L_{\rm s},1)$ this yields to
\be
\Delta_{\pi^0_{1,L_{\rm s}}}(t) =
\Delta_{\pi^0_{L_{\rm s},1}}(t) &\sim& \sum_{{\rm x}} \left [
\left | Q[U]^{-1}_{(0,0),({\rm x},{\rm t});1,L_{\rm s}}  \right | ^2 +
\left | Q[U]^{-1}_{(0,0),({\rm x},{\rm t});L_{\rm s},1}  \right | ^2 \right ]
\ .
\en
For every lattice size we performed 10000 sweeps. Every 100th sweep we
calculated $Q[U]^{-1}$, the correlation function and 
the pion mass separately on every configuration. 
The statistical error was estimated by binning.
The autocorrelation times for the pion mass were of order 
${\cal O}(100)$ sweeps. The maximum value of approximate 500
sweeps was obtained for $\beta=8.0$, $\Omega=32^2 \cdot 4$
and $m=0.01$. 

In fig.\ \ref{KSF4.2} the different correlation functions for the 
$32^2 \cdot 4$ lattice and $(s,s')=(1,L_{\rm s})$ are plotted. From 
this we extracted the pion mass by a least square fit 
by a single cosh. The values are shown in fig.\ \ref{KSF5.2}
together with the masses of the other lattice. One can see that the values of 
the two masses for $m=0.20$ coincide within the statistical errors.
For small values of $m$ finite size effects appear.
An estimate for the value of the critical mass by polynomial interpolation
gives $|m_{\rm c} | \le 0.01$, i.e.\ no deviation of a multiplicative 
renormalization for the mass is observable. Since we are in the weak coupling
region renormalization effects are small anyway. The renormalization of the 
fermion mass can be obtained by comparing 
the data with the results of other simulations
and perturbation theory. The result is 
$\overline{m}=0.90(5) \, m$.

Renormalization effects at $\beta=0.25$ are larger. Therefore
the simulations at this point are a better check for the prediction that 
the fermion mass gets only multiplicative renormalization. As before there 
is no hint for a nonzero critical mass (fig.\ \ref{KSF6.2}). Moreover 
$m/g$ is small enough for a comparison with perturbation theory. The results
are in agreement with theory and lead to
$\overline{m}=0.60(2) \, m$.

\section*{Acknowledgments}

I thank Istv\'{a}n Montvay and Karl Jansen 
for helpful discussions and comments. 
Especially I benefit from Elfie's lectures.

\end{document}